\documentclass{pasj00}

\begin{document}
\SetRunningHead{Uchiyama et al.}{Restoration of Suzaku Astrometry}
\Received{2007/06/14}
\Accepted{2007/09/18}

\title{Restoring the Suzaku Source Position Accuracy \\ and 
Point-Spread Function}
\author{Yasunobu \textsc{Uchiyama},\altaffilmark{1} 
Yoshitomo \textsc{Maeda},\altaffilmark{1} 
Masatoshi \textsc{Ebara},\altaffilmark{1} \\
Ryuichi \textsc{Fujimoto},\altaffilmark{2} 
Yoshitaka \textsc{Ishisaki},\altaffilmark{3}
Manabu  \textsc{Ishida},\altaffilmark{1} \\
Ryo \textsc{Iizuka},\altaffilmark{1,4} 
Masayoshi \textsc{Ushio},\altaffilmark{1} 
Hirohiko \textsc{Inoue},\altaffilmark{1} \\
Shunsaku \textsc{Okada},\altaffilmark{1}
Hideyuki \textsc{Mori},\altaffilmark{5}
and 
Masanobu \textsc{Ozaki}\altaffilmark{1}
}
\altaffiltext{1}{Department of High Energy Astrophysics,
  Institute of Space and Astronautical Science (ISAS), \\
  Japan Aerospace Exploration
  Agency (JAXA), 3-1-1 Yoshinodai, Sagamihara, 229-8510}
\altaffiltext{2}{Graduate School of Natural Science and Technology, Kanazawa University, \\
 Kakuma-machi, Kanazawa, 920-1192}
\altaffiltext{3}{Department of Physics, Tokyo Metropolitan University, \\
 1-1 Minami-Osawa, Hachioji, Tokyo 192-0397}
\altaffiltext{4}{Nishi-Harima Astronomical Observatory, \\
407-2 Nishikawauchi, Sayo-cho, Sayo-gun, Hyogo, 679-5313}
\altaffiltext{5}{Department of Physics, Kyoto University,
Sakyo-ku, Kyoto 606-8502}

\maketitle

\begin{abstract}
We present an empirical correction of  sky coordinates of X-ray photons 
obtained with the X-ray Imaging Spectrometer (XIS) aboard the Suzaku satellite to improve the source position accuracy 
and restore the point-spread function (PSF). 
The XIS images are known to have an uncertainty in position of 
up to $1\arcmin$, 
and to show considerable degradations of the PSF. 
These problems are caused by a drifting of the satellite attitude 
due to thermal distortion of the side panel \#7, 
where the attitude control system is mounted.
We found that the position error averaged over a 
pointing observation can be largely reduced by 
using the relation between the deviation of the source position 
 in the DETX direction and the ecliptic latitude of 
 the pointing target. We parameterized the wobbling of the source position 
synchronized with the 96-minute satellite orbital period with 
 temperatures of onboard radiators 
and elapsed time since the night-day transition of the spacecraft. 
We developed software, aeattcor, to correct the image drift 
using these parameters, and applied it to 27 point-source images 
taken  in the Suzaku initial science-operation phase.
We show that the radius of the 90\% error circle 
of the source position was reduced to $19\arcsec$ and the PSF was sharpened.  
These improvements have enhanced the scientific capability of the Suzaku XIS.
\end{abstract}
 
\section{Introduction}

The X-ray Observatory Suzaku \citep{mitsuda07}, developed
jointly by Japan and the US, has 
two active  scientific payload, XIS (X-ray Imaging Spectrometer: \cite{koyama07}) 
and HXD (Hard X-ray Detector: \cite{takahashi07}).
XIS consists of four X-ray imaging CCD cameras covering a field-of-view of 
 $\timeform{17'.8} \times \timeform{17'.8}$; 
 three are front-illuminated (0.4--12 keV), 
 and one is a back-illuminated CCD (0.2--12 keV).
The XISs  are located on the focal plane of the X-ray Telescopes, 
XRTs \citep{serlemitsos07}.
Each X-ray photon recorded by the XIS can be localized with the CCD 
native pixels, 
each of which has dimensions of  $1.04\arcsec \times 1.04\arcsec$. 
While the PSF of the XRTs has a broad distribution with
a half-power diameter (HPD) of $\sim 2\arcmin$, 
it has a sharp core with a width of $\sim 10\arcsec$.
Thus,  
the sky positions of  point-like sources with good photon statistics 
($\gtrsim  10^4\ \rm photons$) are in principle determined with 
a precision of better than  $\sim 10\arcsec$, 
 if we know the instantaneous pointing direction of the XRT perfectly.
The positional information is crucial to search for 
counterparts in other wavelengths [e.g., a new supersoft source discovered 
by Suzaku; see \citet{takei07}]
and also to construct source catalogs [e.g., \citet{Ueda01} and \citet{Ueda05}
in the case of ASCA]. 

The celestial coordinates of each photon detected with the Suzaku XRT-XIS 
are calculated from the position coordinates of the CCD pixels defined
on the XRT focal plane and the instantaneous pointing direction of the XRT.
The arcsecond localizations  of 
optical guide stars by a pair of star trackers (STTs)
 primarily determine the absolute pointing direction of the XRT.
While guide stars are not available, 
a set of gyroscopes takes a role to determine the pointing direction.
However, over the course of in-orbit calibrations of the Suzaku XRT, 
it was revealed that the calculated positions of X-ray sources
have an error of $1\arcmin$ \citep{serlemitsos07}, which is 
significantly larger than that expected in the pre-flight specifications, $\sim 10\arcsec$.
The deviation of the calculated source position from the true position 
is time-dependent and synchronized 
with the orbital motion of the spacecraft.
Its behavior differs from one observation to another. 
The error of the source position can be as large as $1\arcmin$, and therefore 
the astrometric accuracy of the XIS images has been taken to 
be  $\sim 1\arcmin$ so far. 

\begin{figure}[htb]
\begin{center}
\FigureFile(8cm,8cm){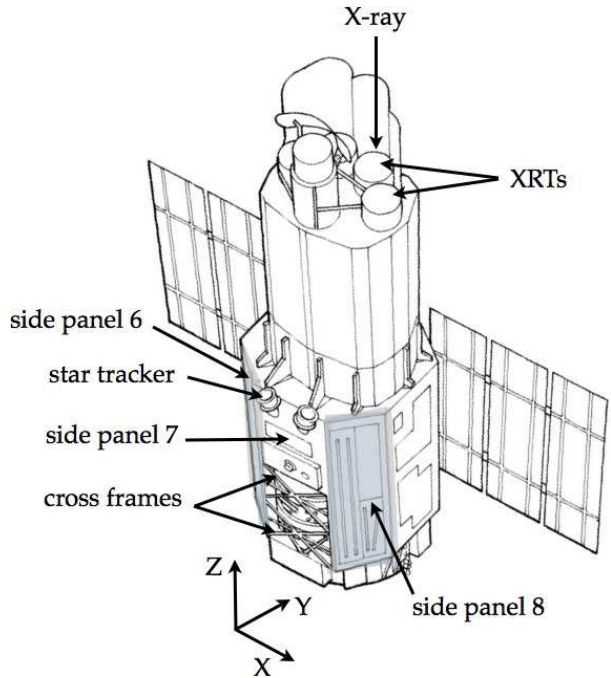}
\end{center}
\caption{Schematic view of the Suzaku satellite. Some key components 
in this work are indicated by arrows. Side panels \#6 and \#8 are indicated 
by grey shaded areas.
The spacecraft (S/C) XYZ 
directions are shown. The DETX direction coincides with the S/C $+$X 
direction, while DETY  is defined by the S/C $-$Y direction. }
 \label{fig:suzaku}
\end{figure}

The cause of this problem has been identified with a thermal distortion of 
the satellite chassis, particularly the cross frames on the side panel \#7
(see figure~\ref{fig:suzaku}), as briefly stated in  \citet{serlemitsos07}.
This panel is situated on the anti-Sun side of the spacecraft,
 where the dewar of the X-ray Spectrometer (XRS: \cite{kelley07}) is placed. 
To expose the XRS dewar to space, the lower half of the panel \#7 is skeletonized 
by cross frames  made  of aluminum 
[see figure~\ref{fig:suzaku} and also \citet{mitsuda07}]. 
The cross frames  suffer from thermal distortion due mainly to Earth albedo 
illumination.
Specifically, side panel \#7 is forced to be tilted  in the DETY direction 
as the cross frames expand, 
since the upper end of the panel is fixed at the side panels \#6 and \#8.
Since the STT and gyroscopes are co-located on side panel \#7, 
it causes a time-dependent misalignment 
between the XRT-XIS system and the attitude system, and 
consequently drifts  the XRT pointing in the DETY direction. 
Also, an imbalance of thermal expansion between the 
cross frames induces rotation around the Y-axis
(perpendicular to the side panel \#7), 
resulting in drifts  in the DETX direction. 

We have investigated the characteristics of drift of the X-ray images 
to restore the pointing accuracy as close to the pre-flight specification as possible.
 Suzaku does not have onboard sensors for 
the temperatures of the cross frames, which are considered to be the most relevant parameters to the thermal distortion.
We therefore have searched for relevant key parameters within the limited housekeeping data available on ground. 

In the present work, 
we parameterize the drifts of the XRT pointing direction 
against the attitude control system with the spacecraft temperatures, 
the satellite orbital phase, and  the average pointing 
direction relative to the ecliptic plane, and 
found  a method to correct the position accuracy and the PSF in the 
observed XIS image using these parameters.
We applied the method  to point-source images 
taken  in the initial scientific operation 
phase and derived the position accuracy currently achievable. 
The outline of this paper is as follows.
We describe the data sets and basic reduction in section~\ref{sec:obs}. 
The parameterization of drifts of the XRT pointing is presented in 
section~\ref{sec:analysis},
which is followed by a brief description of  new software, 
{\tt aeattcor},  developed for the position-error  correction in 
section~\ref{sec:soft}.
We present the improvement on the PSF and the position accuracy
in section~\ref{sec:results}. A summary is in section~\ref{sec:summary}.

\section{Observations}
\label{sec:obs}

We analyzed XIS data acquired during the Suzaku 
initial science-operation phase for the present work. 
 We selected data of point sources whose celestial 
 coordinates were cataloged with a precision of better than a few 
 arcseconds,  their X-ray fluxes were 
higher than $ 0.5\ \rm counts\ s^{-1}\ chip^{-1}$, 
but less than $ 100\ \rm counts\ s^{-1}\ chip^{-1}$ 
because a too-high count-rate caused a significant pileup at the peak of the PSF 
and degraded  the precision of the position determination.

Table 1 lists the selected X-ray sources, consisting mainly of 
active galactic nuclei and X-ray binaries, used in this work.
(For multi-epoch monitoring observations, we 
 used only the data of  the two latest epochs 
in order not to enhance statistical weights on these sources.)
We retrieved the Suzaku XIS data with version 1.2
pipeline processing (V1.2), and utilized cleaned event data. We analyzed the XIS images 
using sky coordinates, 
X/Y  columns (+X is directed to west and +Y is north) in the event files, 
 which were calculated from the detector coordinates (DETX/Y) of the
 CCD pixels and the satellite attitude monitored by the STTs and 
 gyroscopes \citep{ishisaki07}.
It is worth emphasizing that prior to the pipeline
processing V1.0, the attitude determination 
with STTs and gyroscopes  on ground had not  been applied.
We present results obtained from XIS-0, one of the three front-side illuminated 
CCDs, because the XIS-0  has the best PSF suited for our purpose.  
The drifts of all four XIS images due to the drifts of the satellite 
attitude were found to be almost identical.
The results from XIS-0 are equally applicable to the other XIS data.

\section{Analysis}
\label{sec:analysis}

\subsection{Position Drift of Suzaku XIS Image}
\label{sec:wobbling}

In order to track a drift of a source position in the  sky coordinates (X/Y),
we determined the quasi-instantaneous peak of the sky 
image for every 200 second time bin by a two-dimensional Lorentzian fitting. 
The deviation of the peak position from the {\it expected} position 
($\Delta X$, $\Delta Y$) was computed at each time bin.
The expected pixel position is (768.5, 768.5) 
if the  source was aimed to be placed at the XIS nominal position.
They were then converted into the detector coordinate 
system, ($\Delta X_{\rm det}$, $\Delta Y_{\rm det}$), using the 
following transformation:
\begin{eqnarray}
\Delta X_{\rm det} &=& \Delta X \cos\theta + \Delta Y \sin\theta  \\
\Delta Y_{\rm det} &=& -\Delta X \sin\theta + \Delta Y \cos\theta,  
\end{eqnarray}
where $\theta$ is the average roll angle, PA\_NOM, stored in the 
event file headers. 
A similar method was adopted in \citet{Gotthelf00} for correcting 
the ASCA position accuracy.

\begin{figure}[htb]
\begin{center}
\FigureFile(8cm,8cm){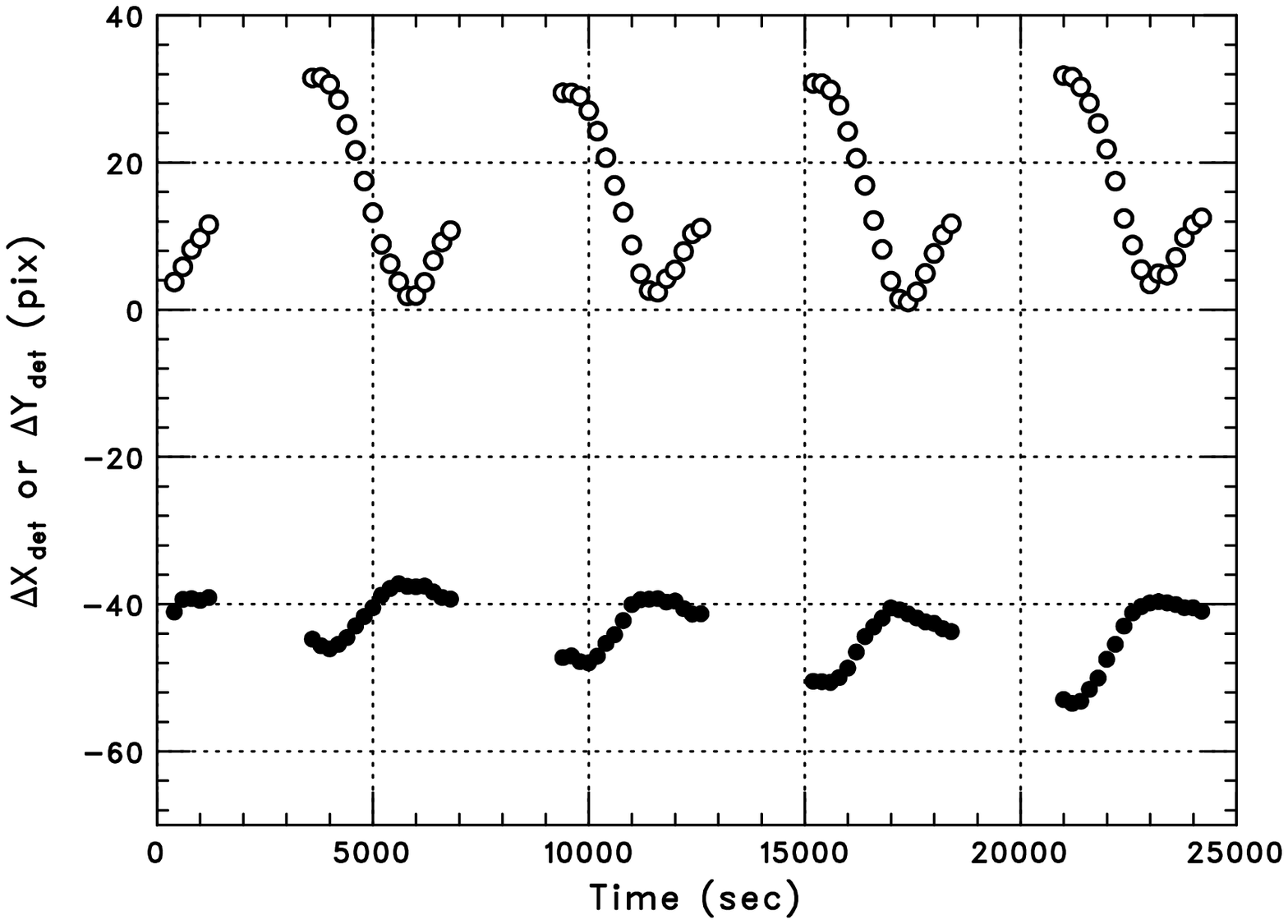}
\FigureFile(8cm,8cm){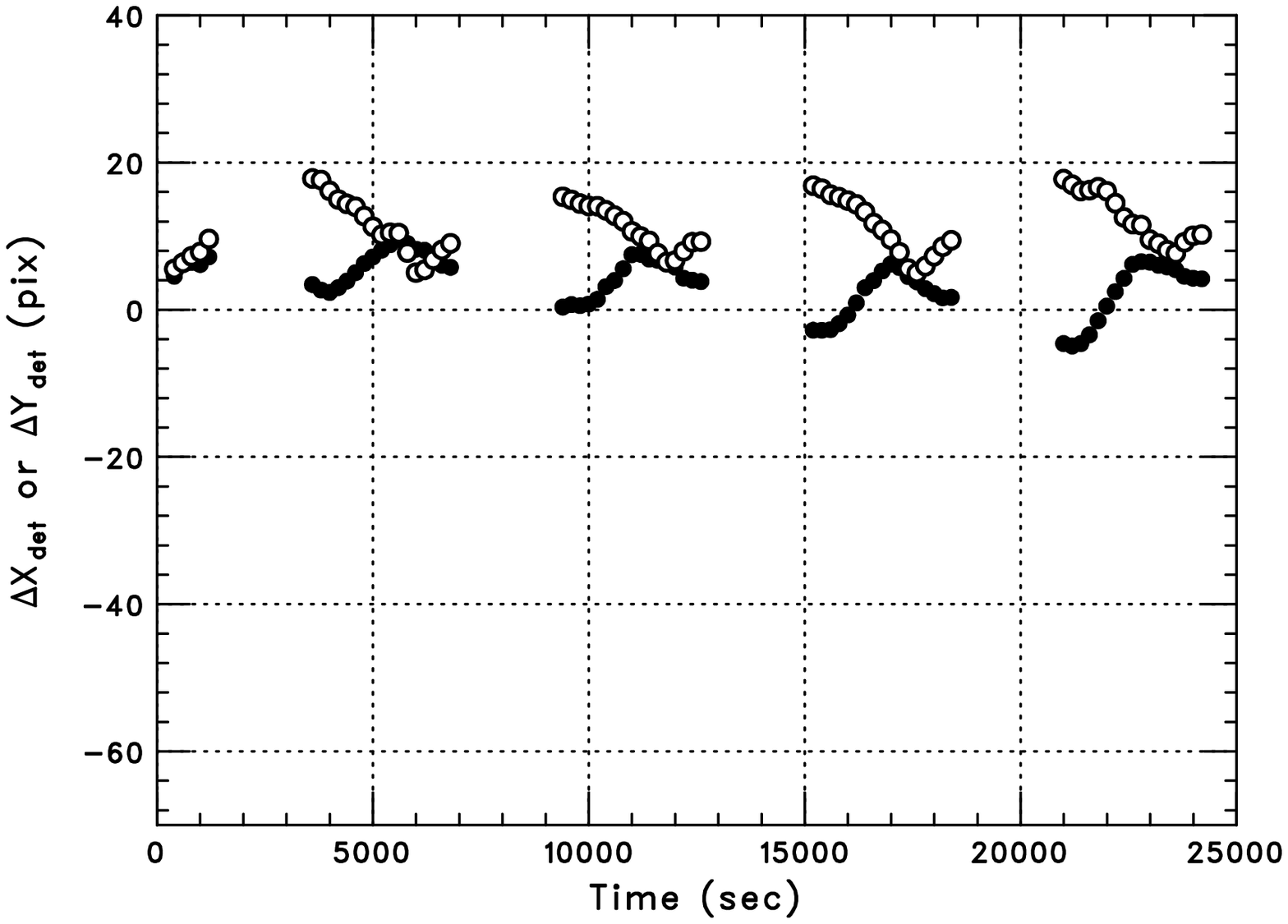}
\end{center}
\caption{(Top) Time sequence of 
$\Delta X_{\rm det}$ (filled circles) and  $\Delta Y_{\rm det}$ (open circles)
in the case of  Her~X-1.  
One pixel  corresponds to $\simeq 1.04 \arcsec$.
The origin of the source position 
is its nominal position, at which a source is expected to fall. 
(Bottom) Same as the top panel, but after the attitude-error correction 
(see subsection~\ref{sec:correlation}). }
 \label{fig:wobbling}
\end{figure}

To illustrate the characteristics  of  the drift of the XIS images, 
the time history of a source position obtained from data of 
Her~X-1 (seq\# 100035010) is shown in 
figure~\ref{fig:wobbling}.
It can be seen that 
the source position oscillates with amplitudes of $10\arcsec$--$30\arcsec$ 
 in both $\Delta X_{\rm det}$ and  $\Delta Y_{\rm det}$ 
synchronized with the satellite orbital motion of a 96-min period.
Furthermore, the center of oscillation has an offset
toward the $-\Delta X_{\rm det}$ 
direction by $\sim 40\arcsec$, which causes a large astrometric error in 
the Suzaku XRT-XIS system. 

We found that the position drift of the Suzaku XIS  image 
can be well described by a wobbling motion synchronized with 
the satellite orbital motion and an offset which changes slowly 
in a much longer timescale than the orbital period during a fixed pointing 
observation.
The offset appears to differ from one pointing to another.
We derived the offset in both  the $X_{\rm det}$ and $Y_{\rm det}$ directions 
for each pointing 
by fitting the time-integrated XIS image with a two-dimensional Gaussian
function. 
We denote the derived offsets as $\langle \Delta X_{\rm det} \rangle$ and 
$\langle \Delta Y_{\rm det} \rangle$, though these quantities are not 
necessarily the average  of the above deviations  
$\Delta X_{\rm det}$ and $\Delta Y_{\rm det}$. 
The offsets would be a good measure of the center of  oscillation. 

\begin{figure*}[htb]
\begin{center}
\FigureFile(8cm,8cm){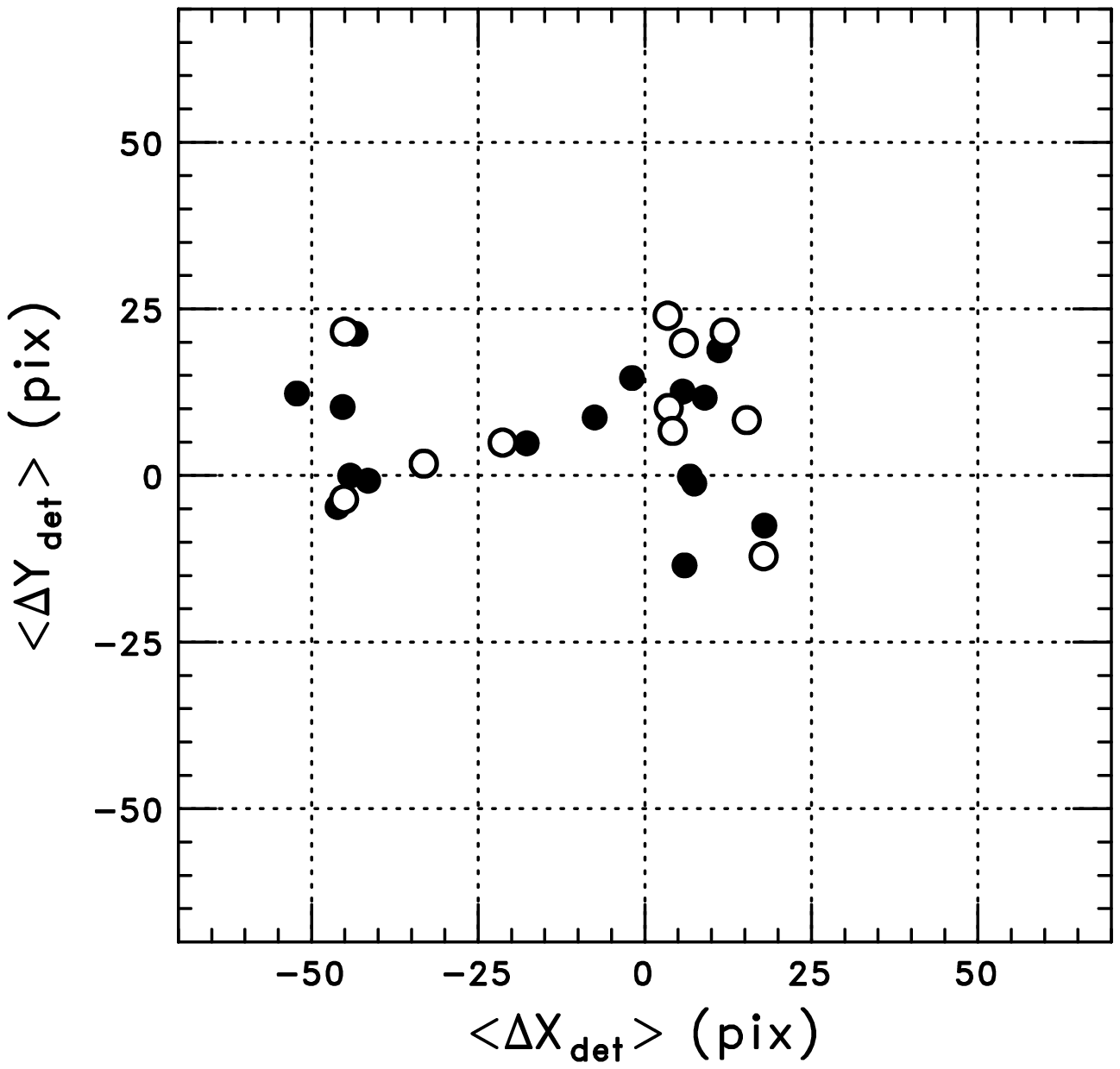}
\FigureFile(8cm,8cm){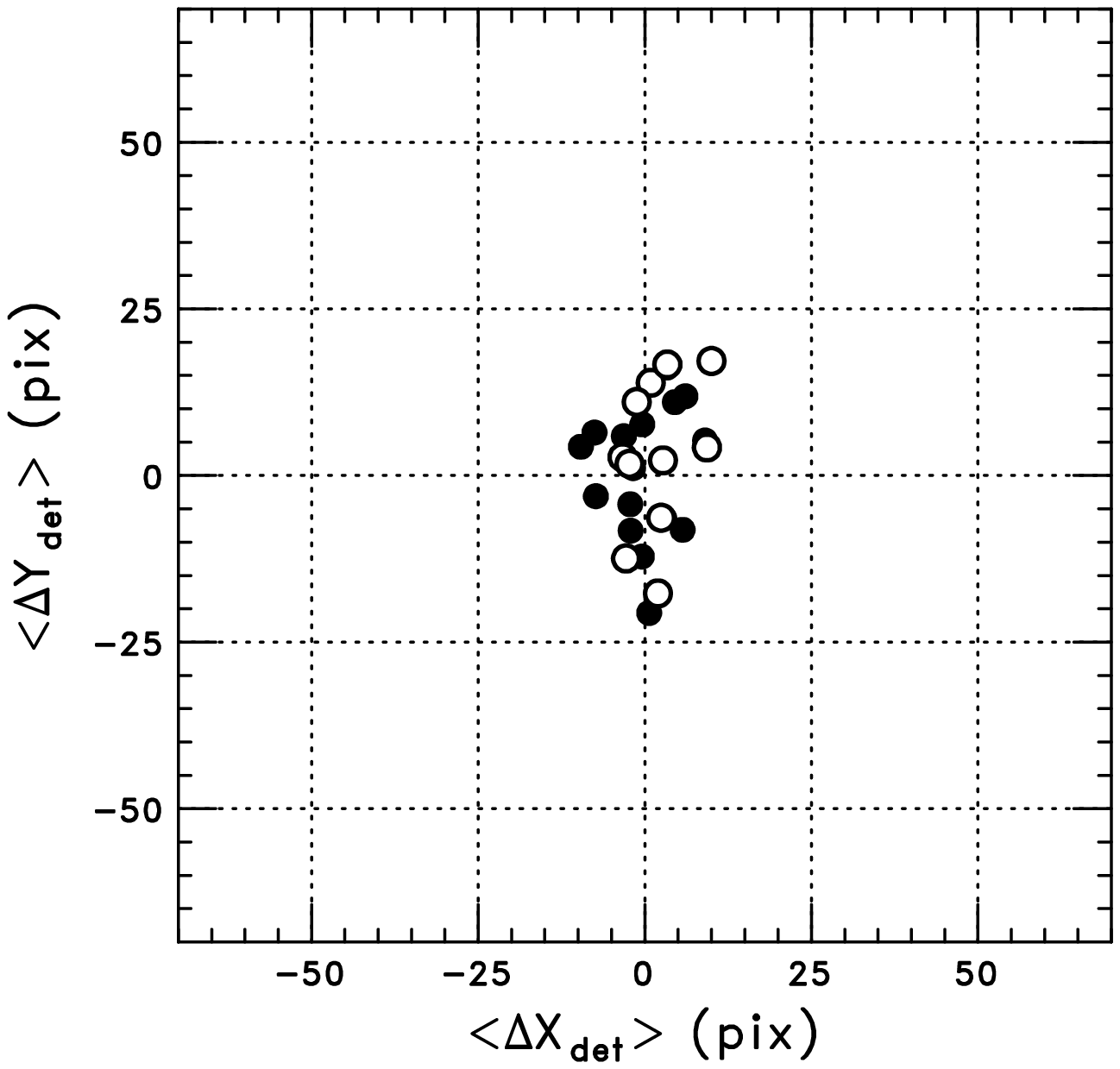}
\end{center}
\caption{(Left) Offsets of XIS source positions from the nominal positions:
$\langle \Delta X_{\rm det} \rangle$ versus  $\langle \Delta Y_{\rm det} \rangle$.
1 pixel $\simeq 1.04 \arcsec$. 
Aim points of the observations are placed at the XIS nominal position 
(filled circles) or at the HXD nominal position (open circles).
(Right) The same plot after the position correction. }
 \label{fig:detxy}
\end{figure*}

In the left panel of figure~\ref{fig:detxy} we show a scatter plot of the offset, 
$\langle \Delta X_{\rm det} \rangle$ versus  $\langle \Delta Y_{\rm det} \rangle$, 
derived for all of our  samples. This plot represents the position accuracy of
the Suzaku XIS image processed by the pipeline V1.2.
We found that the dispersion of the offsets
 is larger in the DETX ($X_{\rm det}$) direction,  $\sim 70\arcsec$. 
The dispersion of the offset in the DETY direction is relatively small  and 
the distribution tends to be positive. 

\subsection{Parametrization of Drift}
\label{sec:correlation}

We searched for related parameters to the offset and the wobbling
motion of the X-ray image to recover the image that should be obtained
in a condition free from the drift of the satellite attitude.
We here assume that the long-timescale offset is constant during 
a fixed pointing observation.

\subsubsection{Long-timescale offset}
\begin{figure}[htb]
\begin{center}
\FigureFile(8cm,8cm){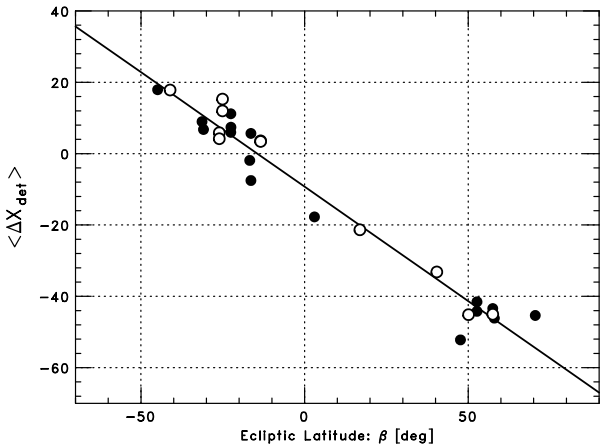}
\end{center}
\caption{Correlation between $\langle \Delta X_{\rm det} \rangle$ and 
the ecliptic latitude $\beta_{\rm ecl}$ of the sources.}
 \label{fig:beta}
\end{figure}

First, we  found that the offset in  DETX, 
$\langle \Delta X_{\rm det} \rangle$, 
has a clear correlation with the ecliptic latitude of the source,  $\beta_{\rm ecl}$.
In figure~\ref{fig:beta} we present the distribution of 
$\langle \Delta X_{\rm det} \rangle$ of each observation
against  $\beta_{\rm ecl}$.
A linear relation between the DETX offset 
$\langle \Delta X_{\rm det} \rangle$ and 
the ecliptic latitude $\beta_{\rm ecl}$  is clearly seen.
The ecliptic latitude
 $\beta_{\rm ecl}$ represents the angle of the XRT pointing direction
 from  the ecliptic plane,  and characterizes  the way of illumination by 
 the bright Earth. 
The strong correlation suggests that 
the geometry between the Sun, Earth, and the satellite, 
plays an essential roll in determining the long-timescale offset of the 
XRT pointing in the DETX direction.
The configuration of  side panel \#7 relative to the ecliptic plane 
is considered to somehow control the thermal distortion of the cross frames. 

The following linear relation is found to reasonably fit these data:
\begin{equation}
\label{eq:DETX}
\langle \Delta X_{\rm det} \rangle (\beta_{\rm ecl}) 
= -38.5 \left(\frac{\beta_{\rm ecl}}{60 \ \rm deg}\right) - 9.25 \ \ {\rm pixels}.
\end{equation}
This relation gives a  DETX offset for 
a given observation using the ecliptic latitude of the target. 
It can be seen that $\langle \Delta X_{\rm det} \rangle \simeq 0$ for 
$\beta_{\rm ecl} \simeq \timeform{-14D}$.
This is simply because the nominal position has been calibrated  by 
 MCG$-$6$-$30-15 (seq\# 100004010), which has an ecliptic latitude of 
$\beta_{\rm ecl} \simeq \timeform{-22.D5}$.

\subsubsection{Wobbling with satellite orbital period}
We next explored  the wobbling motion synchronized with 
the orbital motion  of the satellite. 
After inspecting  correlations with various  parameters stored in
the satellite housekeeping data,
we arrived at  two parameters related with the wobbling motion. 
One is a  difference of  temperature between radiators at the side panels \#8  and 
\#6, {\tt T\_86} (in units of Kelvin), 
  defined as 
 \begin{equation}
\label{eq:HK1}
\tt T\_86 \equiv  HK\_XIS\_RAD8\_T1\_CAL - HK\_XIS\_RAD6\_T1\_CAL, 
\end{equation}
where {\tt HK\_XIS\_RAD8\_T1\_CAL} and  {\tt HK\_XIS\_RAD6\_T1\_CAL}
are temperatures of radiators at the side panel \#8 and \#6, respectively.
The newly introduced parameter is found to be a good
 indicator of the wobbling motion in the DETX direction.
The temperature difference  {\tt T\_86}  represents a temperature gradient 
in  side panel \#7 in the DETX direction.
The other parameter is 
the elapsed time after a night-day transition, 
{\tt T\_DY} (in units of second), defined by 
\begin{equation}
\label{eq:HK2}
\tt 
T\_DY \equiv  \left\{
\begin{array}{ll}
\tt  T\_DY\_NT &  \mbox{for day-time,} \\
\tt  -TN\_DY\_NT & \mbox{for night-time,}
 \end{array}
\right.
\end{equation}
where day-time means a period in which Sun light irradiates the 
satellite.
We make use of the parameter {\tt  T\_DY} to parameterize the 
wobbling motion in the DETY direction because it is supposed to be 
coupled with thermal cycling of the orbiting spacecraft.

\begin{figure}[htbp]
\begin{center}
\FigureFile(8cm,8cm){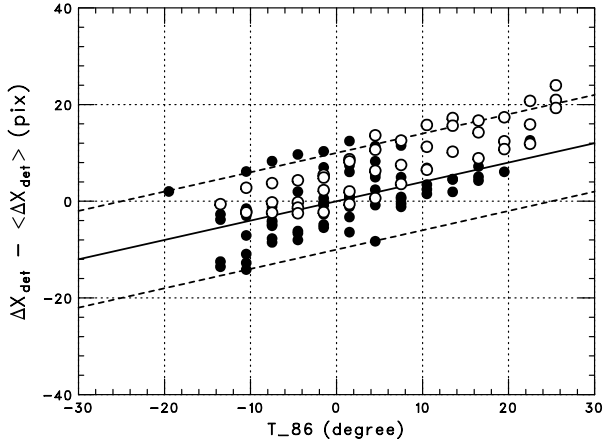}
\end{center}
\caption{Correlation between {\tt T\_86} and 
$\Delta X_{\rm det}-\langle \Delta X_{\rm det} \rangle $, 
derived by tracking the wobbling motion of 
bright sources. The filled points correspond to the XIS nominal cases 
and the open circles to the HXD nominal. 
The solid line represents 
equation~(\ref{eq:T86}), and the dashed lines show a range of scatter in 
$\langle \Delta X_{\rm det} \rangle $. }
\label{fig:T86}
\end{figure}

Figure~\ref{fig:T86} shows a correlation between {\tt T\_86} and $\Delta X_{\rm det}$, 
after removing the offset of $\langle \Delta X_{\rm det} \rangle$.
In the plot, we utilized only bright sources with a flux higher than 
$2\ \rm  counts\ s^{-1}$ among the samples in table 1, because 
 the source position has to be determined for each 200-sec bin.
For a fixed value of {\tt T\_86} in a given observation, 
$\Delta X_{\rm det}$ was found to have a scatter of about 10 pixels. 
We averaged $\Delta X_{\rm det}$ for each temperature bin over 
each observation. 
There is a tendency  that the source position 
deviates from the nominal position to the $+$DETX direction 
as the temperature difference {\tt T\_86} increases. 
We therefore parametrized $\Delta X_{\rm det}$ (pix)  
by the combination of a linear function of {\tt T\_86}
and $\langle \Delta X_{\rm det} \rangle (\beta_{\rm ecl})$ in 
equation~(\ref{eq:DETX}) as 
\begin{equation}
\label{eq:T86}
\Delta X_{\rm det}  =
\langle \Delta X_{\rm det} \rangle (\beta_{\rm ecl}) + 0.4\times {\tt T\_86}.
\end{equation}

\begin{figure}[htbp]
\begin{center}
\FigureFile(8cm,8cm){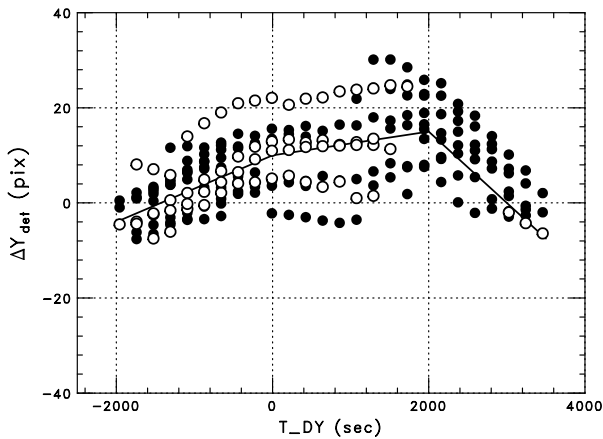}
\end{center}
\caption{Correlation between {\tt T\_DY} and $\Delta Y_{\rm det}$. 
The filled points correspond to the XIS nominal cases 
and the open circles to the HXD nominal. 
The function defined by equation~(\ref{eq:DETY}) is drawn.}
\label{fig:TDY}
\end{figure}

In figure~\ref{fig:TDY} we demonstrate a correlation seen between 
{\tt T\_DY} and $\Delta Y_{\rm det}$. It appears that $\Delta Y_{\rm det}$ 
increases as {\tt T\_DY} increase for ${\tt T\_DY} \leq  2000$ and 
then starts to fall. We approximated the relation between 
$\Delta Y_{\rm det}$ (pix) and  {\tt T\_DY} with a broken line: 
\begin{equation}
\label{eq:DETY}
\Delta Y_{\rm det}
= \left\{ 
\begin{array}{ll}
10 +  0.007\times {\tt T\_DY}
 &  \mbox{for}\ {\tt T\_DY} \leq 0,\\
10 +  0.0025\times {\tt T\_DY}
 &  \mbox{for}\ 0< {\tt T\_DY} \leq 2000, \\
45 -  0.015\times {\tt T\_DY} 
 &  \mbox{for}\ 2000 < {\tt T\_DY}.
\end{array}
\right.
\end{equation}

These empirical relations give an instantaneous  deviation from the nominal position
at each event time for a given observation, thus allowing corrections
in the DETY direction.
Note however that 
the absence of a direct  measurement  of the cross frame temperatures 
makes the corrections difficult. 

\section{Software Implementation}
\label{sec:soft}

We have developed a program,  {\tt aeattcor}, to correct 
the parameters of the satellite attitude using the parameterization of drift of 
the XIS image described above.

The program first calculates the expected deviation of the XRT 
pointing direction from that aimed at each event time using 
 equations~(\ref{eq:T86}) and (\ref{eq:DETY}).
The deviation corresponds to the error of the satellite attitude caused 
 by any misalignment between the XRT-XIS system and the attitude control 
 system.
The program calculates Euler angles of the
satellite attitude for the XRT-XIS system from those monitored on the
attitude control system, and then updates
the attitude file in which the Euler angles are stored.
If one simply
converts detector coordinates (DETX, DETY) to sky coordinates (X,Y)
using the updated attitude file, the corrected sky coordinates (X,Y) can be obtained.
A detailed description of the conversion scheme is given by \citet{ishisaki07}.

The program is now included in FTOOLS, the official analysis 
software released form the Suzaku Guest Observer Facility, and will be 
applied to the XIS event files released through the pipeline processing 
V2.0 or the later versions.

\section{Demonstration}
\label{sec:results}

In this section we demonstrate the developed program, 
{\tt aeattcor}, in application to the XIS images of sample sources.
We show how degree the program improves the PSF and 
the position accuracy.

\subsection{Restored Image and PSF}

\begin{figure}[htbp]
\begin{center}
\FigureFile(8cm,8cm){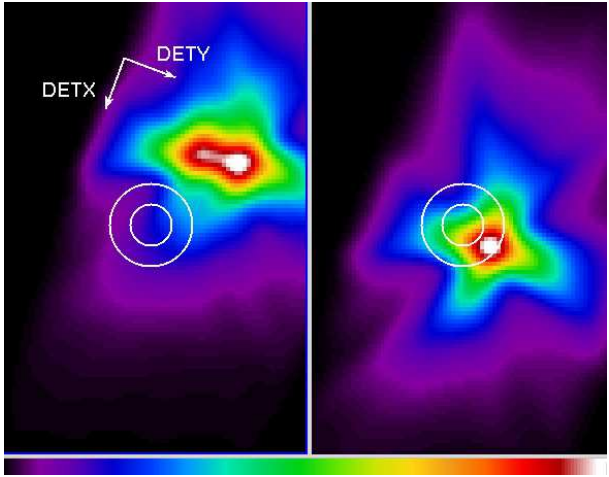}
\end{center}
\caption{Time-integrated XIS-0 images made with (X, Y) columns 
of Her~X-1 are displayed before (left) and after (right) the position correction. 
Counts per pixel normalized at the peak are shown on a linear color scale. 
North (+Y) is up and west (+X) is to the right. 
The double circles with a radius of $10\arcsec$ and $20\arcsec$ 
show the nominal (expected) position of the source.
The source peak in the 
corrected image in the right panel is closer to the nominal position.
Furthermore, a fake double-core of the original image is clearly converged. }
\label{fig:image}
\end{figure}

The left panel of figure~\ref{fig:image} shows
time-integrated XIS-0 image of Her~X-1 
 before the attitude-error correction, 
as extracted  from the V1.2 event files. 
The image before the attitude-error correction
looks as if there are two point-like sources separated by 30\arcsec .
It is because the position of Her~X-1 is 
 oscillated with the satellite orbital period, 
and the source visibility is limited in selected orbital phases for 
Earth occultation (see figure~\ref{fig:wobbling}).  

The right panel of figure~\ref{fig:image} is an  image after the 
attitude-error correction.
One can realize that the deviation of the peak  from the nominal position 
is improved, and the double-peak profile of the image before the 
correction disappears in the corrected image.
In figure~\ref{fig:wobbling}, we present a time sequence of the source position 
after the correction, where 
the large amplitude  in the DETY is reduced.

To quantify the improvement of  the PSF core,
we  introduce a new parameter, 
``core sharpness" $S$, defined as  the ratio of the averaged count per pixel 
within an inner circle of 10-pix radius to that within an outer 
 annulus between 10-pix and 20-pix radii.
Here we determined the source centroid by fitting with 
a 2-dimentional Gaussian function.
If the PSF is perfectly flat within 20 pixels from the center, 
the core sharpness is 1.0.
For the instantaneous PSF free from the time-variable pointing direction, 
we expect $S\simeq 2$. 
We note that the ``core sharpness" depends on each  XIS 
 because each XRT has a slightly different PSF. We here present the  results of XIS-0.

\begin{figure}[htbp]
\begin{center}
\FigureFile(8cm,8cm){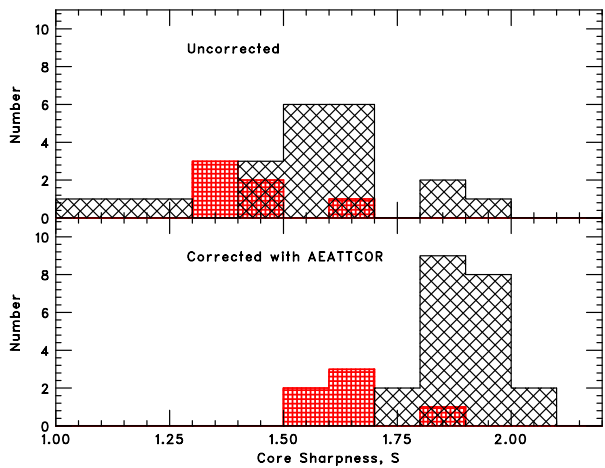}
\end{center}
\caption{Distributions of core sharpness $S$ for XIS-0. 
The top panel shows the distribution of $S$ before the correction, 
while the bottom panel shows that after it. 
$S=1$ corresponds to a flat PSF in the central 20-pix and $S=2$ 
to the undistorted PSF core. Results for bright sources exceeding 50 counts 
exposure$^{-1}$, 
which are likely to be affected by photon pile-up, are separately shown in red. }
\label{fig:sharp}
\end{figure}

Figure~\ref{fig:sharp} shows the distributions of $S$ for the 
sources listed in table 1, before and after the attitude-error correction.
The sources with a flux higher than 
 $50\ \rm counts\ exposure^{-1}\ chip^{-1}$ 
are shown separately with red histograms, because those bright sources 
are expected to severely suffer from pileup effects. 
It is clear that 
 the  PSFs  of the corrected image are sharpened and the core sharpness 
 becomes close to 2, 
the value expected for the PSF free from the attitude error.

\subsection{Improved Position Accuracy}
\label{sec:correction}

We determined  the offset of the source centroid 
from the expected position for all sample sources  before/after 
the attitude-error correction, in order to evaluate the improvement 
on the source position accuracy.
The distributions of these offsets 
($\langle \Delta X_{\rm det} \rangle ,  \langle \Delta Y_{\rm det} \rangle $)
before/after the attitude-error correction are plotted separately in 
figure~\ref{fig:detxy}. 
The improvement of the position accuracy is clearly seen
from the difference of the two distributions.

\begin{figure}[htpb]
\begin{center}
\FigureFile(8cm,8cm){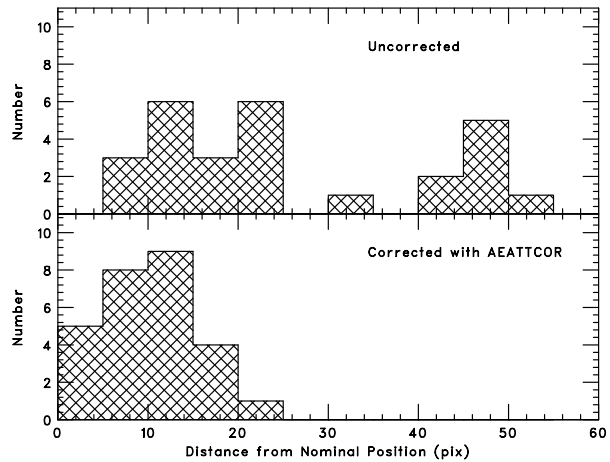}
\end{center}
\caption{Distributions of the source distance $\Delta r$ from the nominal position.
The top panel shows the $\Delta r$ distribution before 
the attitude-error  correction, while the bottom panel shows that after correction.}
\label{fig:hist}
\end{figure}

To quantify the position accuracy, 
figure~\ref{fig:hist} shows the distributions of 
$\Delta r \equiv 
\sqrt{\langle \Delta X_{\rm det} \rangle^2 +  \langle \Delta Y_{\rm det} \rangle ^2}$, 
i.e., the distance between the source centroid and the expected position, 
separately before/after the attitude-error correction. 
Without the attitude-error correction, 
the deviation from the nominal position has 
 a 90\% error circle of  $\Delta r \simeq 50\arcsec$ radius.
The  {\tt aeattcor}  restores  the pointing accuracy 
 to $\Delta r \leq 19\arcsec$ (a 90\% error circle).
The improvement is brought mainly by the successful 
parameterization of the large DETX offsets. 
 The residual position error after the attitude-error correction 
is somewhat worse compared with the preflight specification,  
$\sim 10\arcsec$, and the position accuracy of ASCA, $\simeq 12\arcsec$ 
 \citep{Gotthelf00}.
Given a lack of direct measurements of 
the temperatures of the cross frames, which are the most relevant parameters,
we do not expect that further improvements can be easily made. 
We will monitor the  Suzaku pointing accuracy to check its validity 
over the course of the mission. 

\section{Summary}
\label{sec:summary}

We have developed 
an empirical method to restore the position accuracy and the PSF in the 
observed Suzaku XIS image. 
We analyzed 27 pointing observations of point-like sources 
using the XIS-0 data of 
the pipeline processing V1.2 to investigate the characteristics of the drift of 
the XIS image which degrades the position accuracy and the PSF. 
We found that the deviation of the source centroid from the expected 
position, averaged over one pointing observation, can be well characterized  by 
 the relation between the deviation 
in the DETX direction and the ecliptic latitude of the pointing target.
It is also found that 
 the oscillation of the source position 
synchronized with the satellite orbital period can be 
parameterized to some extent by 
the temperatures of the onboard radiators 
and the elapsed time since the night-day transition of the satellite. 
Based on the parameterization of the drifts of the XIS image, we 
have implemented new software, {\tt aeattcor}, to correct the XRT 
pointing direction with respect to the attitude control system at each 
event time. 
By applying the attitude-error corrections to 
the XIS data of our  sample, 
we demonstrated that the distorted core of PSF can be sharpened and 
showed that a  90\% error circle  of the sky position of 
a point-like source is reduced to $19\arcsec$ in radius. 
The results presented in this paper describe the current accuracy of performing 
astrometry with the Suzaku  XIS. 
\\

We thank the anonymous referee for useful comments and suggestions, 
which improved the paper. 
We are very grateful to Takashi Usui, Akira Okamoto, Kazuyo Mizushima,
Tetsu Saitoh, Kazunori Shouji, Takayuki Tohma (NEC Co.) who gave us 
valuable comments on the thermal distortions of the Suzaku satellite.

\begin{longtable}{lcrrrrrrrrr}\label{tab:list}
\caption{List of Suzaku Point Sources Used in This Study}
\hline
  & Suzaku  & R.A. & Decl. & $\beta_{\rm ecl}$ & Aim\footnotemark[$*$] & $\theta$\footnotemark[$\dagger$]
  & $\Delta X_{\rm cor}$ & $\Delta Y_{\rm cor}$ 
  & $\Delta X$ & $\Delta Y$\\
Object & Sequence \# & (J2000.0) & (J2000.0)  & (deg) & & (deg) 
 & (pix) & (pix) & (pix) & (pix) \\
\hline
\endhead
\hline
\endfoot
\hline
\multicolumn{11}{l}{\hbox to 0pt{\parbox{180mm}{\footnotesize
\smallskip
Notes.---Units of right ascension are hours, minutes, and seconds, and units of
declination are degrees, arcminutes, and arcseconds. 
$\Delta X_{\rm cor}$ and $\Delta Y_{\rm cor}$ are pixel deviations of the 
source centroid from the nominal position in the corrected X/Y image, while 
$\Delta X$ and $\Delta Y$ are for the uncorrected image.\\
\footnotemark[$*$] Aim point of the observation. 
XIS: the target is placed at the nominal optical axis of the XIS. 
HXD: the target is placed at the nominal optical axis of the HXD, which 
is shifted in DETX by $\timeform{-3.5'}$ from the XIS optical axis.  \\
\footnotemark[$\dagger$] Roll angle in the observation, stored 
as {\tt PA\_NOM} in the header of the event FITS files.\\
\footnotemark[$\ddagger$] These sources have $>50\ \rm counts\ exposure^{-1}$ 
(0.4--10 keV) and consequently show photon pile-up in the central regions 
(within about 20 pixels from the center) of the PSF. \\ 
}}}
\endlastfoot
MCG$-$6$-$30-15 & 100004010 & 13 35 53.80 & $-$34 17 44.0 & $-$22.5 
& XIS & 296.0 & 13.4 & $-$0.3 & 21.8 & $-$1.8 \\ 
Cen A & 100005010 & 13 25 27.60 & $-$43  01 08.8 & $-$31.3 
& XIS & 303.9  & 6.2 & 4.6 & 14.7 & $-$0.9 \\ 
1A 0535+26 & 100021010 &  05 38 54.57 & 26 18 56.8 & 3.0 
& XIS & 83.6& 2.3 & $-$7.6 & $-$6.8 & $-$17.1 \\ 
NGC 2110 & 100024010 &  05 52 11.40 & $-$07 27 22.0 & $-$30.9 
& XIS & 103.4  & 8.5 & $-$0.1 & $-$1.4 & 6.6 \\ 
NGC 3516 & 100031010 & 11  06 47.50 & 72 34 07.0 & 58.0 
& XIS & 148.9  & $-$0.6 & 9.9 & 41.9 & $-$19.8 \\ 
Her X-1\footnotemark[$\ddagger$] & 100035010 & 16 57 49.83 & 35 20 32.6 & 57.5 
& XIS & 249.6  & 14.0 & $-$8.7 & 35.1 & 33.3 \\ 
Mkn 3 & 100040010 &  06 15 36.30 & 71  02 15.0 & 47.6 
& XIS & 71.8 & $-$7.1 & $-$7.8 & $-$28.0 & $-$45.8 \\ 
Her X-1\footnotemark[$\ddagger$] & 101001010 & 16 57 49.83 & 35 20 32.6 & 57.5 
& HXD & 67.9  & $-$12.6 & 6.1 & $-$37.0 & $-$33.6 \\ 
PKS 2155$-$304 & 101006010 & 21 58 52.00 & $-$30 13 32.0 & $-$16.8 
& XIS & 57.9 & $-$6.7 & 0.5 & $-$13.4 & 6.2 \\ 
AE Aquarii & 400001010 & 20 40 09.16 & $-$00 52 15.1 & 16.9 
& HXD & 264.9 & 3.0 & 3.1 & 6.8 & 20.8 \\ 
GX 349+2\footnotemark[$\ddagger$] & 400003010 & 17  05 44.50 & $-$36 25 23.0 & $-$13.4 
& HXD & 85.7  & $-$2.1 & 2.9 & $-$9.8 & 4.3 \\ 
GX 349+2\footnotemark[$\ddagger$] & 400003020 & 17  05 44.50 & $-$36 25 23.0 & $-$13.4 
& HXD & 78.9  & $-$15.7 & 6.5 & $-$22.9 & 8.0 \\ 
SS Cyg & 400006010 & 21 42 42.80 & 43 35 09.9 & 52.7 
& XIS & 276.6  & $-$4.5 & 1.7 & $-$5.5 & 41.2 \\ 
SS Cyg & 400007010 & 21 42 42.80 & 43 35 09.9 & 52.7 
& XIS & 256.9 & 1.5 & 1.5 & 10.0 & 43.1 \\ 
X1630$-$472\footnotemark[$\ddagger$] & 400010050 & 16 34 01.10 & $-$47 23 34.4 & $-$25.1 
& HXD & 107.7  & $-$6.8 & 7.6 & $-$12.6 & 12.1 \\ 
X1630$-$472\footnotemark[$\ddagger$] & 400010060 & 16 34 01.10 & $-$47 23 34.4 & $-$25.1 
& HXD & 120.3  & $-$19.9 & 0.0 & $-$24.6 & $-$0.5 \\ 
4U 1626$-$67 & 400015010 & 16 32 16.80 & $-$67 27 43.0 & $-$44.9 
& XIS & 103.0  & 11.9 & 2.3 & 3.3 & 19.1 \\ 
CH Cyg & 400016020 & 19 24 33.07 & 50 14 29.1 & 70.5 
& XIS & 185.8  & $-$8.5 & $-$6.1 & 46.2 & $-$5.7 \\ 
3C 120 & 700001020 &  04 33 11.10 &  05 21 15.0 & $-$16.4 
& XIS & 237.5  & 6.9 & $-$9.7 & 7.6 & $-$11.6 \\ 
3C 120 & 700001030 &  04 33 11.10 &  05 21 15.0 & $-$16.4 
& XIS & 262.5 & 7.4 & 6.6 & 9.6 & 6.3 \\ 
MCG$-$5$-$23-16 & 700002010 &  09 47 40.10 & $-$30 56 56.0 & $-$41.1 
& HXD & 118.4 & 14.6 & 10.1 & 2.2 & 21.4 \\ 
NGC 4051 & 700004010 & 12  03 09.60 & 44 31 53.0 & 40.4 
& HXD & 135.3  & 2.7 & 6.2 & 22.3 & $-$24.6 \\ 
NGC 2992 & 700005020 &  09 45 42.00 & $-$14 19 35.0 & $-$26.1 
& HXD & 88.8  & $-$11.1 & $-$1.0 & $-$19.8 & 6.3 \\ 
NGC 2992 & 700005030 &  09 45 42.00 & $-$14 19 35.0 & $-$26.1 
& HXD & 128.5  & 0.1 & $-$2.9 & $-$7.8 & $-$0.9 \\ 
MCG$-$6$-$30-15 & 700007020 & 13 35 53.80 & $-$34 17 44.0 & $-$22.5
& XIS & 102.4  & 20.0 & 5.1 & 11.9 & 8.7 \\ 
MCG$-$6$-$30-15 & 700007030 & 13 35 53.80 & $-$34 17 44.0 & $-$22.5 
& XIS & 102.4 & 5.7 & 4.3 & $-$0.4 & 7.5 \\ 
0836+710 & 700010010 &  08 41 24.30 & 70 53 42.0 & 50.1 
& HXD & 294.8 & $-$12.5 & $-$2.7 & $-$22.2 & 39.5 \\ 
\end{longtable}

\end{document}